\documentclass[copyright,creativecommons]{eptcs}
\providecommand{\event}{iFMCloud 2016}

\usepackage{cleveref}
\crefname{section}{Sect.}{Sect.}
\Crefname{section}{Section}{Sections}
\crefname{figure}{Fig.}{Fig.}
\Crefname{figure}{Figure}{Figures}
\crefname{definition}{Def.}{Def.}

\newcommand{\cpp}{\textsc{C++}}

\usepackage{graphicx}
\usepackage{caption,subcaption}
\newlength\heightfiga\newlength\heightcapa
\newlength\heightfigb\newlength\heightcapb
\newlength\heightfigc\newlength\heightcapc
\newlength\heightfig

\usepackage{tikz}
\usetikzlibrary{arrows,positioning,patterns,decorations.pathreplacing}

\usepackage{syntax} 
\usepackage{amsmath,mathtools,amsthm}

\newcommand{\rel}{\sqsubseteq}
\newcommand{\constr}[2]{\ensuremath{#1 \rel #2}}

\newcommand{\nil}{\ensuremath{\mathsf{nil}}}

\newcommand{\true}{\ensuremath{\mathsf{true}}}
\newcommand{\false}{\ensuremath{\mathsf{false}}}

\newcommand{\cspws}{\textsc{CSP-WS}}

\newcommand{\con}{\ensuremath{\mathcal{C}}}

\usepackage{etoolbox}

\newcommand{\el}[3]{%
    \mathsf{#1}\ifx$#2$\else(#2)\fi{:}~{#3}
}

\newcounter{listcount}\newcounter{totalcount}%
\newcommand{\tuple}[1]{%
  \setcounter{totalcount}{0}
  \renewcommand*{\do}[1]{\stepcounter{totalcount}}
  \docsvlist{#1}
  \setcounter{listcount}{0}
  \renewcommand*{\do}[1]{
    \stepcounter{listcount}
    ##1\ifnum\value{listcount}<\value{totalcount}\,\fi
  }
  \ensuremath{(\docsvlist{#1})}
}
\newcommand{\tlist}[2]{%
  \setcounter{totalcount}{0}
  \renewcommand*{\do}[1]{\stepcounter{totalcount}}
  \docsvlist{#1}
  \setcounter{listcount}{0}
  \renewcommand*{\do}[1]{
    \stepcounter{listcount}
    ##1\ifnum\value{listcount}<\value{totalcount},\else\fi
  }
  \ensuremath{[\docsvlist{#1}\ifx$#2$\else\,| #2\fi]}
}
\newcommand{\record}[2]{%
  \setcounter{totalcount}{0}
  \renewcommand*{\do}[1]{\stepcounter{totalcount}}
  \docsvlist{#1}
  \setcounter{listcount}{0}
  \renewcommand*{\do}[1]{
    \stepcounter{listcount}
    ##1\ifnum\value{listcount}<\value{totalcount}, \else\fi
  }
  \ensuremath{\{\docsvlist{#1}\ifx$#2$\else\,| #2\fi\}}
}
\newcommand{\choice}[2]{%
  \setcounter{totalcount}{0}
  \renewcommand*{\do}[1]{\stepcounter{totalcount}}
  \docsvlist{#1}
  \setcounter{listcount}{0}
  \renewcommand*{\do}[1]{
    \stepcounter{listcount}
    ##1\ifnum\value{listcount}<\value{totalcount},\else\fi
  }
  \ensuremath{{(}{:}\docsvlist{#1}\ifx$#2$\else\,| #2\fi{:}{)}}
}
\newcommand{\switch}[1]{%
  \setcounter{totalcount}{0}
  \renewcommand*{\do}[1]{\stepcounter{totalcount}}
  \docsvlist{#1}
  \setcounter{listcount}{0}
  \renewcommand*{\do}[1]{
    \stepcounter{listcount}
    ##1\ifnum\value{listcount}<\value{totalcount},\else\fi
  }
  \ensuremath{\langle\docsvlist{#1}\rangle}
}

\newcommand{\eel}[3]{%
    \mathsf{#1}\ifx$#2$\else(#2)\fi{:}~{#3}
}

\newcommand{\etuple}[1]{%
  \setcounter{totalcount}{0}
  \renewcommand*{\do}[1]{\stepcounter{totalcount}}
  \docsvlist{#1}
  \setcounter{listcount}{0}
  \renewcommand*{\do}[1]{
    \stepcounter{listcount}
    ##1\ifnum\value{listcount}<\value{totalcount}\,\fi
  }
  \ensuremath{(\docsvlist{#1})}
}
\newcommand{\etlist}[2]{%
  \setcounter{totalcount}{0}
  \renewcommand*{\do}[1]{\stepcounter{totalcount}}
  \docsvlist{#1}
  \setcounter{listcount}{0}
  \renewcommand*{\do}[1]{
    \stepcounter{listcount}
    ##1\ifnum\value{listcount}<\value{totalcount},\else\fi
  }
  \ensuremath{[\docsvlist{#1}\ifx$#2$\else\,| #2\fi]}
}
\newcommand{\erecord}[2]{%
  \setcounter{totalcount}{0}
  \renewcommand*{\do}[1]{\stepcounter{totalcount}}
  \docsvlist{#1}
  \setcounter{listcount}{0}
  \renewcommand*{\do}[1]{
    \stepcounter{listcount}
    ##1\ifnum\value{listcount}<\value{totalcount}, \else\fi
  }
  \ensuremath{\{\docsvlist{#1}\ifx$#2$\else\,| #2\fi\}}
}
\newcommand{\echoice}[2]{%
  \setcounter{totalcount}{0}
  \renewcommand*{\do}[1]{\stepcounter{totalcount}}
  \docsvlist{#1}
  \setcounter{listcount}{0}
  \renewcommand*{\do}[1]{
    \stepcounter{listcount}
    ##1\ifnum\value{listcount}<\value{totalcount},\else\fi
  }
  \ensuremath{{(}{:}\docsvlist{#1}\ifx$#2$\else\,| #2\fi{:}{)}}
}
\newcommand{\eswitch}[1]{%
  \setcounter{totalcount}{0}
  \renewcommand*{\do}[1]{\stepcounter{totalcount}}
  \docsvlist{#1}
  \setcounter{listcount}{0}
  \renewcommand*{\do}[1]{
    \stepcounter{listcount}
    ##1\ifnum\value{listcount}<\value{totalcount},\else\fi
  }
  \ensuremath{\langle\docsvlist{#1}\rangle}
}

\newcommand{\eupvar}[1]{\textsf{\textbf{#1}}}
\newcommand{\edownvar}[1]{\textsf{\textbf{#1}}}

\usepackage{fancyvrb}

\theoremstyle{definition}
\newtheorem{definition}{Definition}

\title{Configuring Cloud-Service Interfaces Using Flow Inheritance}
\author{Pavel Zaichenkov \qquad Olga Tveretina \qquad Alex Shafarenko
\institute{University of Hertfordshire, United Kingdom}
\email{\{p.zaichenkov,o.tveretina,a.shafarenko\}@ctca.eu}
}

\def\titlerunning{Configuring Cloud-Service Interfaces Using Flow Inheritance}
\def\authorrunning{P.~Zaichenkov, O.~Tveretina \& A.~Shafarenko}
\begin{document}
\maketitle

\begin{abstract}
Technologies for composition of loosely-coupled web services in a modular and
flexible way are in high demand today.  On the one hand, the services must be
flexible enough to be reused in a variety of contexts.  On the other hand, they
must be specific enough so that their composition may be provably consistent.
The existing technologies (WS-CDL, WSCI and session types) require
a behavioural contract associated with each service, which is impossible to
derive automatically.  Furthermore, neither technology supports flow
inheritance: a mechanism that automatically and transparently propagates data
through service pipelines.  This paper presents a novel mechanism for automatic
interface configuration of such services.  Instead of checking consistency of
the behavioural contracts, our approach focuses solely on that of data formats
in the presence of subtyping, polymorphism and flow inheritance.  The paper
presents a toolchain that automatically derives service interfaces from the
code and performs interface configuration taking non-local constraints into
account.  Although the configuration mechanism is global, the services are
compiled separately.  As a result, the mechanism does not raise source security
issues despite global service availability in adaptable form.
\end{abstract}

\section{Introduction}

Web services is a technology that facilitates development of large scale
distributed applications.  The application is composed from stand-alone
services provided by various organisations on the web.  However, consistent
composition and reliable coordination of such applications still remain
challenging for the following reasons: 1) the services are provided as loosely
coupled black boxes that only expose their interfaces to the environment; 2)
interacting services are not usually known in advance: web services are
dynamically chosen to fulfil certain roles and are often replaced by services
with a similar functionality; 3) the nature of the service-based application is
decentralised~\cite{sheng2014web}.

Over the last decade, a substantial body of research on communication safety in
web services has been produced.  Typically, service behaviour is defined as
a contract specified via a suitable process algebra.  Various formalisms for
verifying the properties of the consistent composition exist.  For example, in
session types~\cite{dezani2010sessions} a composition is defined as a global
protocol in terms of the interactions that are expected from the protocol
peers, and a set of local protocols, one for each peer, which describe the
global protocol from the viewpoint of an individual peer.  Another formalism is
called interface automata~\cite{hashemian2005graph}.  It provides a method that
constructs complex services with a specific behaviour from simpler services
given semantic properties of the simple services.  These approaches support
guaranteed communication safety, deadlock-freedom and protocol conformance in
service-based applications~\cite{honda2008multiparty}.

\paragraph{Flow inheritance.}Web services are experiencing transition from
batch to stream processing~\cite{akidau2013millwheel}.  Under stream processing
services often form long pipelines, where a service processes only a part of
its input message, with the rest of it bypassed down the pipeline without
modification.  A mechanism that implicitly redirects a part of a message from
input to output of a service is called \emph{flow
inheritance}~\cite{grelck2010asynchronous}.  Although existing technologies for
service composition, such as session types, enhance flexibility and reuse of
services with subtyping and polymorphism, they lack analysis and configuration
capacity for flow inheritance.  Providing support for flow inheritance in
stateful services is not a trivial problem.  A complete transition system,
which specifies the behavioural protocol, needs to be analysed to trace the
path of the inherited data.  In this paper we focus only on stateless services
that do not have an internal storage and which produce output messages in
response to a single input message.

In~\cite{zaichenkov2016constraint} we presented a formal method for configuring
interfaces in the presence of subtyping, polymorphism and flow inheritance.
The method is based on constraint satisfaction and Boolean satisfiability
(SAT)\@.  In our approach, the interface of a single service is defined in
generic form that is compatible with a variety of contexts.  The main features
of the generic interface are 1) free variables that implement parametric
polymorphism by instantiation to particular context-specific values; 2) Boolean
variables that are used to control the dependencies between any elements of
interface collections in a way similar to intersection
types~\cite{pierce1997intersection} (those map input type to output type for an
overloaded function); and 3) support of flow inheritance using extensible
records and variants, which are implemented in a form of row
polymorphism~\cite{gaster1996polymorphic}, that can be extended with additional
elements that propagate data formats and functionality across the global
communication graph.  We illustrate using a running example introduced
in~\cref{sec:example}.

This paper presents a toolchain that performs a complete interface
configuration.  First, it derives the interfaces from the \cpp{} code of
individual services.  We demonstrate that in our approach the complete
interfaces can automatically be derived from the code.  Next, the toolchain
produces constraints for the interfaces of interconnected services given the
derived interfaces and a specification of the communication graph.  Then
a CSP is solved and the values assigned to the variables are sent back to the
services.  Finally, the services are separately compiled using those values
inserted in the code.  In \cref{sec:implementation} we
demonstrate that this approach does not raise vulnerability issues in
security-critical services, because a service does not require a centralised
compilation (in particular, a service can be separately compiled for each
context and be exported in binary).

\section{Constraint Satisfaction Problem for Web Services}

In this section we briefly introduce a novel Interface Description Language
called Message Definition Language (MDL) as an alternative to Web Service
Description Language (WSDL)\@.  The MDL is defined as a term algebra and is
formally introduced in~\cite{zaichenkov2016constraint}.  Its purpose is to
describe flexible service interfaces.

A message is a collection of data entities, each specified by a corresponding
\emph{term}.  The term is the MDL basic building block.  MDL terms are built
recursively according to the following grammar:
\begin{grammar}
    <term> ::=  <symbol> |  term-variable | <record> | <choice>

    <record> ::= "\{"$[$<element>$[$","<element>$]^* [$"|" term-variable$]]$"\}"

    <choice> ::= "(:"$[$<element>$[$","<element>$]^*[$"|" term-variable$]]$":)"

    <element> ::= <label>"("<bool-expr>"):"<term>
\end{grammar}

Each term is either a \emph{symbol}, a term variable or a recursively-defined
labelled collection of terms.  The symbol represents standard types such as
\texttt{int} or \texttt{string}.  Term variables are used to support parametric
polymorphism in the interfaces similarly to type variables.  A \emph{record} is
an extensible, unordered collection of labelled terms, where \emph{labels} are
arbitrary identifiers, which are unique within a single record.  Records are
subtyped covariantly (a larger record is a subtype).  A \emph{choice} is an
extensible, unordered collection of labelled alternative terms.  The difference
between records and choices is in width subtyping: \emph{choices} are subtyped
contravariantly using set inclusion (a smaller choice is a subtype).  We use
choices to represent polymorphic messages and service interfaces at the top
level (see \cref{sec:example,sec:protocol} for more details).  Records and
choices are defined in \emph{tail form}.  The tail is denoted by a term
variable that represents a term of the same kind as the construct in which it
occurs.  This is a form of row polymorphism~\cite{gaster1996polymorphic}, which
we use to implement flow inheritance.  The data that is matched by the same
variable in input and output interfaces is automatically bypassed to other
services in the pipeline (see~\cref{sec:example} for illustration of flow
inheritance).

Elements of the collection may contain arbitrary Boolean expressions.  They are
used to specify relations between input and output interface elements of
a single service as illustrated in \cref{sec:example}.  The intention is
similar to intersection types, which increase the expressiveness of function
signatures: the type $(a \to c) \land (b \to d)$ maps particular input types to
particular output types.  By analogy, assume that the term
$\record{\el{a}{x}{\texttt{int}}, \el{b}{y}{\texttt{int}}}{}$ defines an input
interface of a service and the term $\record{\el{c}{x}{\texttt{int}}, \el{d}{x
\lor y}{\texttt{int}}}{}$ defines the output interface of the service.  The
Boolean variables $x$ and $y$ declare that the service produces messages with
labels \textsf{c} and \textsf{d} only if a message with the label \textsf{a}
was received as an input, and a message with the label \textsf{d} is produced
if a message with the label \textsf{b} was received.  This mechanism allows to
trace message dependencies over the communication graph and check message
format compatibility.

\cref{def:seniority} introduces the seniority relation $\rel$ on terms for the
purpose of structural subtyping.  It specifies the subsumption relation between
interfaces of service ports connected with a communication channel.  In the
sequel we use $\nil$ to denote the empty record $\{\enskip\}$, which has the
meaning of unit type and represents a message without any data.

\begin{definition}
    The seniority relation on terms is defined as follows:
    \begin{enumerate}
        \itemsep0em
        \item $t \rel \nil$ if $t$ is a symbol or a record;
        \item $t \rel t$;
        \item $t_1 \rel t_2$, if for some $k,m>0$ one of the following holds:
        \begin{enumerate}
            \item $t_1
                = \record{\el{l^1_1}{}{t^1_1},\dots,\el{l^k_1}{}{t^k_1}}{}$ and
                $t_2
                = \record{\el{l^1_2}{}{t^1_2},\dots,\el{l^m_2}{}{t^m_2}}{}$,
                where $k \geq m$ and for each $j \leq m$ there is $ i \leq k$
                such that $l^i_1 = l^j_2$ and $t^i_1 \rel t^j_2$;

            \item $t_1
                = \choice{\el{l^1_1}{}{t^1_1},\dots,\el{l^k_1}{}{t^k_1}}{}$ and
                $t_2
                = \choice{\el{l^1_2}{}{t^1_2},\dots,\el{l^m_2}{}{t^m_2}}{}$,
                where $k \leq m$ and for each $i \leq k$ there is $ j \leq m$
                such that $l^i_1 = l^j_2$ and $t^i_1 \rel t^j_2$.
        \end{enumerate}
    \end{enumerate}
    \label{def:seniority}
\end{definition}

Connecting two service ports with a communication channel gives rise to
a constraint on terms, which may contain Boolean and term variables.
Therefore, the problem can be formulated as a constraint satisfaction problem
for web services:
\begin{definition}[\cspws{}]
\label{def:cspws}
    Given a set of seniority constraints $\con$, a vector of Boolean variables
    $\vec{f} = (f_1,\dots,f_l)$ and a vector of term variables
    $\vec{v}=(v_1,\dots,v_m)$, find a vector of Boolean values $\vec{b}
    = (b_1,\dots,b_l)$ and a vector of terms without variables
    $\vec{t}=(t_1,\dots,t_m)$, such that for each constraint $\constr{t_1}{t_2}
    \in \con$,
    \[
        t_1[\vec{f}/\vec{b}][\vec{v}/\vec{t}] \rel
        t_2[\vec{f}/\vec{b}][\vec{v}/\vec{t}]\text{,}
    \]
    where $t[\vec{p}/\vec{q}]$ denotes a substitution of variables $\vec{p}$
    with values $\vec{q}$ ($|\vec{p}| = |\vec{q}|$).  The tuple $(\vec{b},
    \vec{t})$ is called a solution.
\end{definition}

We designed a fixed-point algorithm that solves the $\cspws{}$, which is
presented in~\cite{zaichenkov2016constraint}.  It has been implemented as
a constraint solver, which is used as part of a toolchain presented in
\cref{sec:protocol}.

\section{Motivating Example}
\label{sec:example}

We use a simple but non-trivial example to illustrate the MDL and configuration
mechanism from \cref{sec:protocol}. The example is known as the
\emph{three-buyer use case} and is often called upon to demonstrate the
capabilities of session types~\cite{honda2008multiparty}.

Three buyers called Alice, Bob and Carol cooperate in order to buy a book from
a Seller.  The service interfaces are defined as the MDL terms, which form
a seniority constraint once the services get connected by a channel.
\cref{fig:three-buyer} depicts composition of the application where Alice is
connected to the Seller only and can interact with Bob and Carol indirectly
using flow inheritance.  $AS, SB, BC, CB, BS, SA$ denote interfaces that are
associated with service output/input ports.  For brevity, we only provide $AS,
SB$ and $BC$ (the rest of the interfaces are defined in the same manner), which
are specified in the following way:

\begin{figure}[t]
    \centering
    \begin{tikzpicture}[thick,node distance=2cm,minimum width=3em]
        \node[draw,rectangle] (a) {Alice};
        \node[draw,rectangle,right=of a] (s) {Seller};
        \node[draw,rectangle,right=of s] (b) {Bob};
        \node[draw,rectangle,right=of b] (c) {Carol};

        :Ж\draw[transform canvas={yshift=0.5ex},->] (a) to node[above,near start] {$AS_{\text{out}}$} node[above,near end] {$AS_{\text{in}}$} (s);
        \draw[transform canvas={yshift=-0.5ex},->,dashed] (s) to node[below,near start] {$SA_{\text{out}}$} node[below,near end] {$SA_{\text{in}}$} (a);

        \draw[transform canvas={yshift=0.5ex},->] (s) to node[above,near start] {$SB_{\text{out}}$} node[above,near end] {$SB_{\text{in}}$} (b);
        \draw[transform canvas={yshift=-0.5ex},->,dashed] (b) to node[below,near start] {$BS_{\text{out}}$} node[below,near end] {$BS_{\text{in}}$} (s);

        \draw[transform canvas={yshift=0.5ex},->] (b) to node[above,near start] {$BC_{\text{out}}$} node[above,near end] {$BC_{\text{in}}$} (c);
        \draw[transform canvas={yshift=-0.5ex},->,dashed] (c) to node[below,near start] {$CB_{\text{out}}$} node[below,near end] {$CB_{\text{in}}$} (b);
    \end{tikzpicture}
    \caption{Service composition in a Three Buyer usecase}
\label{fig:three-buyer}
\end{figure}
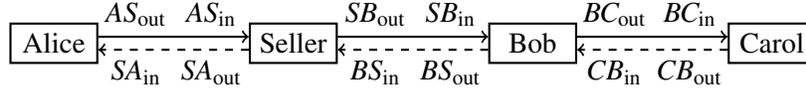

\begin{equation*}
    \scriptsize
    \openup-.5\jot
\begin{aligned}[c]
    AS_{\text{out}} = {(}{:}&\eel{request}{}{\erecord{\eel{title}{}{\edownvar{a}}}{}},\\
                     &\eel{payment}{}{\erecord{\eel{title}{}{\edownvar{a}},
                      \eel{money}{}{\mathtt{int}},
                      \eel{id}{}{\mathtt{int}}}{}},\\
                     &\eel{share}{x}{\erecord{\eel{title}{}{\edownvar{a}},
                      \eel{money}{}{\mathtt{int}}}{}},\\
                             &\eel{suggest}{y}{\erecord{\eel{title}{}{\edownvar{a}}}{}}{:}{)}\\
                      SB_{\text{out}} = {(}{:}&\mathsf{response}:~\{\eel{title}{}{\mathtt{string}},\eel{money}{}{\mathtt{int}}\}~|~\eupvar{b}{:}{)}\\
    BC_{\text{out}} = {(}{:}&\eel{share}{z}{\erecord{\eel{quote}{}{\mathtt{string}}, \eel{money}{}{\mathtt{int}}}{}}~|~\eupvar{c}{:}{)}
\end{aligned}
\left|
\begin{aligned}[c]
    AS_{\text{in}} = {(}{:}&\eel{request}{}{\erecord{\eel{title}{}{\mathtt{string}}}{}},\\
                            &\eel{payment}{}{\erecord{\eel{title}{}{\mathtt{string}},
                            \eel{money}{}{\mathtt{int}}}{}}~|~\eupvar{b}{:}{)}\\
                            SB_{\text{in}} = {(}{:}&\mathsf{share}(z):~\{\eel{quote}{}{\mathtt{string}}, \eel{money}{}{\mathtt{int}}\},\\
                           &\mathsf{response}:~\{\eel{title}{}{\mathtt{string}},\eel{money}{}{\mathtt{int}}\}~|~\eupvar{c}{:}{)}\\
    BC_{\text{in}} = {(}{:}&\mathsf{share}:~\{\eel{quote}{}{\mathtt{string}}, \eel{money}{}{\mathtt{int}}\}{:}{)}
\end{aligned}
\right.
\end{equation*}

$AS_\text{out}$ specifies an output interface of Alice, which declares
functionality and a format of messages sent to the Seller in the following way:

\begin{itemize}
    \itemsep0em
    \item Alice can \textsf{request} a book's price from the Seller by
        providing the \textsf{title} of an arbitrary type (which is specified
        by a term variable $\edownvar{a}$) that the Seller is compatible with.
        On the other hand, the Seller in $AS_\text{in}$ declares that the
        \textsf{title} of type \texttt{string} can only be accepted, which
        means that $\edownvar{a}$ must be instantiated to \texttt{string}.
    \item Furthermore, Alice can provide a \textsf{payment} for a book.  In
        addition to the \textsf{title} and the required amount of
        \textsf{money}, Alice provides her \textsf{id} in the message.
        Although the Seller does not require the \textsf{id}, the
        interconnection is still valid due to the width subtyping supported in
        the MDL\@.
    \item In addition, Alice can offer to \textsf{share} a purchase between
        other customers using flow inheritance.  Although Alice is not
        connected to Bob or Carol and may even not be aware of their presence,
        the inheritance mechanism detects that Alice can send a message with
        ``\textsf{share}'' label to Bob by bypassing the message implicitly
        through the Seller.  In order to enable flow inheritance in Seller's
        service, the mechanism sets the tail variable $\eupvar{b}$ to ${(}{:}
        \eel{share}{}{\{\eel{title}{}{\texttt{string}},
        \eel{money}{}{\texttt{int}}\}}{:}{)}$.  If Bob were unable to accept
        a message with the ``\textsf{share}'' label, the mechanism would
        instantiate $x$ to $\false$, which automatically removes the
        corresponding functionality from the term.
    \item Finally, Alice can \textsf{suggest} a book to other buyers.  However,
        examination of other service interfaces shows that there is no service
        that can receive a message with the label ``\textsf{suggest}''.
        Therefore, a communication error occurs if Alice decides to send the
        message.  To avoid this, the configuration mechanism excludes
        ``\textsf{suggest}'' functionality from Alice's service by setting
        the variable $y$ to $\false$.
\end{itemize}

The proposed configuration mechanism analyses the interfaces of services
Seller, Bob and Carol in the same manner.  The presence of $\eupvar{b}$
variable in both input and output interfaces of Bob enables support of flow
inheritance on the interface level; even though it is defined locally, the
effect of flow inheritance is non-local and potentially involves any number of
actors.  Furthermore, the Boolean variable $z$ behaves as an intersection type:
Bob has ``purchase sharing'' functionality declared as an element
$\el{share}{z}{\{\dots\}}$ in its input interface $SB_\text{in}$ (used by the
Seller).  The element is related to the element $\el{share}{z}{\{\dots\}}$ in
its output interface $BC_\text{out}$ (used by Carol).  The relation declares
that Bob provides Carol with ``sharing'' functionality only if Bob was provided
with the same functionality from the Seller.  In our example, $z$ is $\true$,
because Carol declares that it can receive messages with the label
``\textsf{share}''.  Note that there could be any Boolean formula in place of
$z$, which wires any input and output interfaces of a single service in an
arbitrary way.  The existing interface description languages (WSDL, WS-CDL,
etc.) do not support such interface wiring capabilities.

\section{Interface Configuration Protocol}
\label{sec:protocol}

We developed a complete interface configuration protocol and configuration
toolchain for services coded in \cpp{}.  Choosing \cpp{} allowed us to consider
not only web services, but also performance-critical applications.  On the
other hand, the approach could easily be extended to other languages, which are
commonly used for web service development, such as Java or Python.

We defined a fixed interface format for services.  \cref{fig:service-code}
illustrates the implementation code of the Seller from \cref{sec:example}.
\texttt{request\_1} and \texttt{payment\_1} are processing functions that are
called when an input message arrives; the input data is passed with functions'
arguments.  The processing functions are distinguished from other functions by
their return type \texttt{service}.  On the other hand, output messages are
produced by calling special functions called \emph{salvo}s, which are declared
by the programmer and must have \texttt{salvo} as the return type.  The
processing function sends output messages by calling salvo functions.  The
processing functions and salvos have suffixes \texttt{\_1} or \texttt{\_2} as
part of their names.  The suffixes of the processing functions denote the input
ports that the functions are associated with.  Similarly, the suffix of a salvo
denotes a service output port where the messages are sent to.  Since port
routing is typically specific to the application, association with the ports is
not fixed and can be redefined from the \emph{shell}, which is described in
\cref{sec:shell}.  In \cref{fig:service-code}, error messages are sent to the
second output port, which is considered as an auxiliary one: the application
designer can decide whether they want to process error messages or not (in the
latter case, the port may remain unwired).

\begin{figure}[t]
    \setlength\heightfiga{7.5cm}
    \setlength\heightfigb{4.1cm}
    \setlength\heightcapa{1\baselineskip}
    \setlength\heightcapb{2\baselineskip}
    \setlength\heightcapc{\heightcapa}
    \setlength\heightfigc{\heightfiga+\heightfigb+\heightcapa}
    \setlength\heightfig{\heightfigc+\heightcapc}

    \begin{minipage}[b][\heightfig][t]{0.3\textwidth}\centering
\begin{Verbatim}[fontsize=\ttfamily\fontsize{8}{10}\selectfont]
salvo response_1(string title, int money);
salvo invoice_1(int id);
salvo error_2(string msg);
service request_1(string title) {
    try {
        int price = ...
        response_1(title, price);
    } catch (exception e) {
        error_2(e.what());
    }
}
service payment_1(string title, int money) {
    try {
        int invoice_id = ...
        invoice_1(invoice_id);
    } catch (exception e) {
        error_2(e.what());
    }
}
\end{Verbatim}
    \parbox[b][\heightcapa][t]{1\linewidth}{\subcaption{The code of the Seller service}\label{fig:service-code}}
        {\scriptsize
            \setlength{\abovedisplayskip}{6pt}
            \setlength{\belowdisplayskip}{\abovedisplayskip}
            \setlength{\abovedisplayshortskip}{0pt}
            \setlength{\belowdisplayshortskip}{3pt}
            \begin{align*}
                \text{IN \#1: } &{(:}\el{request}{x}{\record{\el{title}{}{\texttt{string}}}{\edownvar{a}}},\\
                                &\el{payment}{y}{\record{\el{title}{}{\texttt{string}}, \el{money}{}{\texttt{int}}}{\edownvar{b}}}~|~\eupvar{c}{:)}\\
                \text{OUT \#1: } &{(:}\el{response}{x}{\record{\el{title}{}{\texttt{string}},\el{money}{}{\texttt{int}}}{\edownvar{d}}},\\
                                 &\el{invoice}{y}{\record{\el{id}{}{\texttt{int}}}{\edownvar{e}}}~|~\eupvar{c}{:)}\\
                \text{OUT \#2: } &{(:}\el{error}{x \lor y}{\record{\el{msg}{}{\texttt{string}}}{\edownvar{f}}}{:)}
            \end{align*}
        }
        \parbox[b][\heightcapb][t]{1\linewidth}{\subcaption{One input and two
        output interfaces derived from the Seller service's code.  Additional
        constraints $\edownvar{a} \rel \edownvar{d}$,  $\edownvar{a} \rel
        \edownvar{f}$,  $\edownvar{b} \rel \edownvar{e}$, $\edownvar{b} \rel
        \edownvar{f}$ must be generated}\label{fig:interface-terms}}
    \end{minipage}\hfill
    \begin{minipage}[b][\heightfig][t]{0.55\textwidth}
    \centering
\begin{Verbatim}[fontsize=\ttfamily\fontsize{8}{10}\selectfont]
#if defined(BV_x)
salvo response_1(string title, int money TV_d_decl);
#endif
#if defined(BV_y)
salvo invoice_1(int id TV_e_decl);
#endif
#if defined(BV_x) || defined(BV_y)
salvo error_2(string msg TV_f_decl);
#endif
#ifdef BV_x
service request_1(string title TV_a) {
    try {
        int price = ...
        response_1(title, price TV_d_use);
    } catch (exception e) {
        error_2(e.what() TV_f_use);
    }
}
#endif
#ifdef BV_y
service payment_1(string title, int money TV_b) {
    try {
        int invoice_id = ...
        invoice_1(invoice_id TV_e_use);
    } catch (exception e) {
        error_2(e.what()  TV_f_use);
    }
}
#endif
\end{Verbatim}
\parbox[b][\heightcapc][t]{1.\linewidth}{\subcaption{The code for the Seller
        service augmented with preprocessor's directives.  \texttt{BV\_}\dots{}
        and \texttt{TV\_}\dots{} are macro names}\label{fig:annotated-service}}
    \end{minipage}%
    \caption{Illustration of the configuration mechanism using the Seller
    service's code.  Every term/Boolean variable in the interface has
    a corresponding macro definition in the augmented code}
    \label{protocol-example}
\end{figure}

\cref{fig:interface-terms} demonstrates the interfaces that the toolchain
derives by analysing the source file.  We define a fixed interface format for
services as a choice-of-records term.  Labels in the choice term of the input
interface match processing function names tagged with corresponding labels.  As
a result, a message is structured as a labelled data record, where the label
identifies the function that processes the message.  The name of a salvo
corresponds to one of the labels in the output choice term.  The compatibility
of two services connected by a channel is defined by the seniority relation.

The Boolean variables preserve the relation between choice variants in input
and output interfaces: \textsf{response} is present in the
output interface \#1 only if \textsf{request} is present in
the input one; similarly, \textsf{error} is present in the output interface \#2
only if \textsf{request} or \textsf{payment} is present in the input interface.
The intention is to forbid the service from accepting input messages with
certain tags if the messages that the corresponding functions produce cannot be
accepted by the consumer.

In \cref{fig:annotated-service} the Seller service is augmented with macros,
which gets instantiated with particular values after the CSP has been solved.
\texttt{BV\_x} and \texttt{BV\_y} are macros that correspond to Boolean
variables $x$ and $y$.  Depending on instantiations of $x$ and $y$, unnecessary
processing functions and salvos can be removed from the service.

The tail variables are added to records and choices in the interfaces in
\cref{fig:interface-terms}  for flow inheritance support.  For example, the
function \texttt{request\_1} may receive extra arguments that it does not
require itself, but the consumers of output messages (i.e.~the messages with
labels \textsf{response} and \textsf{error} in our example) demand.  Such
arguments are matched by $\edownvar{a}$ in the input interface and are
propagated further in variables $\edownvar{d}$ and $\edownvar{f}$.  The
constraints $\edownvar{a} \rel \edownvar{d}$ and $\edownvar{a} \rel
\edownvar{f}$ specify the relation between the propagated data: the data that
the function \texttt{request\_1} produces must be compatible with the data that
the function receives with its arguments.

In the service code we add macros directives (prefixed with \texttt{TV\_} in
\cref{fig:annotated-service}) to support flow inheritance.  For example, assume
that the \cspws{} solver concludes that the service that consumes a message
with tag \textsf{response} requires information about an author of the
requested book and the producer of the function \texttt{request\_1} may
actually provide this information.  Then the solution is $\edownvar{a}
= \edownvar{d} = \record{\el{author}{}{\texttt{string}}}{}$, and \texttt{TV\_a}
and \texttt{TV\_d\_use} are defined as ``\texttt{, string author}'' and ``\texttt{,
author}'' respectively.

\label{sec:shell}

\paragraph{The Shell.}  So far the flexibility of our approach is impeded by
fixed labels in service interfaces.  The problem is that the corresponding
names in the MDL interfaces of the connected services must be identical for
a safe communication.  For example, a term that specifies an output message
$\record{\el{user\_id}{}{\texttt{int}}}{}$ and the one that specifies an input
message $\record{\el{user}{}{\texttt{int}}}{}$ are not compatible because of
the label mismatch.  The services cannot be decontextualised because a service
developer is forced to modify the services according to the context where the
service is used.

\begin{figure}[t]
\centering
    \begin{minipage}[t]{0.4\textwidth}
        \centering
        \includegraphics[width=2.5in]{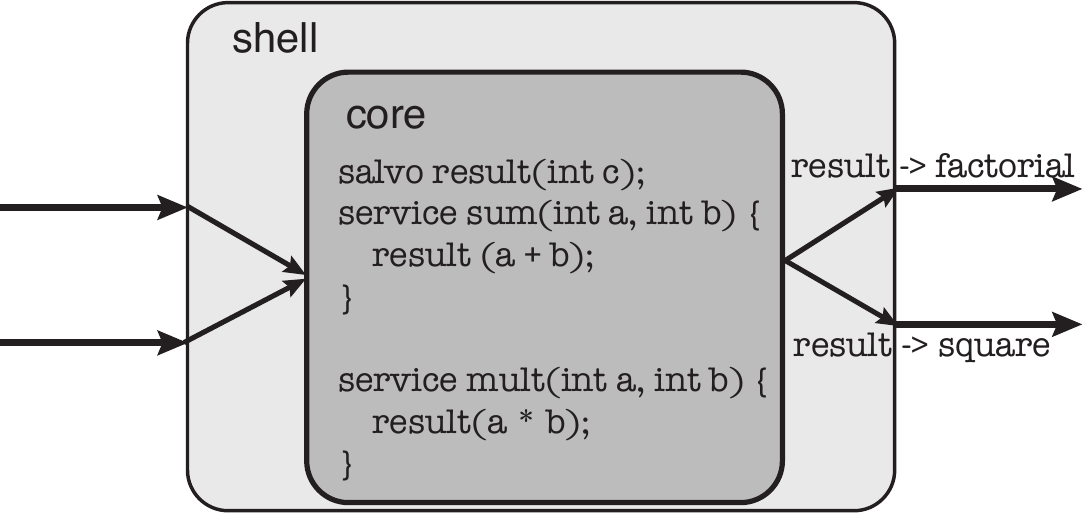}
        \captionof{figure}{The core and the shell of the service}
        \label{fig:core-shell}
    \end{minipage}
    \hspace{2em}
    \begin{minipage}[t]{0.5\textwidth}
    \centering
\tikzset{%
        auto,
        >=latex',
        capt/.style={%
            minimum width=1.3cm,
            node distance=1mm},
        vertex/.style={%
               rectangle,
               rounded corners,
               draw=black,
               thick,
               minimum width=1.3cm,
           node distance=1mm}
    }
    \scalebox{.7}{
    \begin{tikzpicture}[auto,font=\footnotesize]
        \node[capt] (alice) {Alice};
        \node[capt,left=of alice] (seller) {Seller};
        \node[capt,right=of alice] (bob) {Bob};
        \node[capt,right=of bob] (carol) {Carol};

        \node[vertex,below=of alice] (v1a) {\texttt{A.cpp}};
        \node[vertex,left=of v1a] (v1s) {\texttt{S.cpp}};
        \node[left=1mm of v1s] (v1caption) {Source code};
        \node[vertex,right=of v1a] (v1b) {\texttt{B.cpp}};
        \node[vertex,right=of v1b] (v1c) {\texttt{C.cpp}};

        \node[vertex,below=3mm of v1a] (v2a) {\texttt{A.cpp}$\mathtt{^*}$};
        \node[vertex,left=of v2a] (v2s) {\texttt{S.cpp}$\mathtt{^*}$};
        \node[left=1mm of v2s] (v2caption) {Augmentation with macros};
        \node[vertex,right=of v2a] (v2b) {\texttt{B.cpp}$\mathtt{^*}$};
        \node[vertex,right=of v2b] (v2c) {\texttt{C.cpp}$\mathtt{^*}$};

        \node[vertex,below=3mm of v2a] (v3a) {\texttt{A.int}};
        \node[vertex,left=of v3a] (v3s) {\texttt{S.int}};
        \node[left=1mm of v3s] (v3caption) {Interface derivation};
        \node[vertex,right=of v3a] (v3b) {\texttt{B.int}};
        \node[vertex,right=of v3b] (v3c) {\texttt{C.int}};
        \node[vertex,right=of v3c,text width=1.3cm] (topology) {\texttt{topology}};

        \node[vertex,below=3mm of v3b] (constrs) {\texttt{seniority constraints}};
        \node[left=1.58cm of constrs] (constrcaption) {Constraint generation};

        \node[vertex,below=3mm of constrs] (solution) {\texttt{\cspws{}}};
        \node[left=2.95cm of solution] (solutioncaption) {CSP solution};

        \node[vertex,below=3mm of solution] (v4b) {\texttt{B.hpp}};
        \node[vertex,left=of v4b] (v4a) {\texttt{A.hpp}};
        \node[vertex,left=of v4a] (v4s) {\texttt{S.hpp}};
        \node[left=1mm of v4s] (v4caption) {Header file generation};
        \node[vertex,right=of v4b] (v4c) {\texttt{C.hpp}};

        \node[circle,minimum size=.01mm,inner sep=0pt,outer sep=0pt,right=2cm of v2c] (dummy) {};

        \node[vertex,below=3mm of v4a] (v5a) {\texttt{A.lib}};
        \node[vertex,left=of v5a] (v5s) {\texttt{S.lib}};
        \node[left=1mm of v5s] (v5caption) {Library compilation};
        \node[vertex,right=of v5a] (v5b) {\texttt{B.lib}};
        \node[vertex,right=of v5b] (v5c) {\texttt{C.lib}};

        \draw[->,thick] (v1s) -- (v2s);
        \draw[->,thick] (v1a) -- (v2a);
        \draw[->,thick] (v1b) -- (v2b);
        \draw[->,thick] (v1c) -- (v2c);

        \draw[->,thick] (v2s) -- (v3s);
        \draw[->,thick] (v2a) -- (v3a);
        \draw[->,thick] (v2b) -- (v3b);
        \draw[->,thick] (v2c) -- (v3c);

        \draw[->,thick] (v3s.south) -- (constrs);
        \draw[->,thick] (v3a.south) -- (constrs);
        \draw[->,thick] (v3b.south) -- (constrs);
        \draw[->,thick] (v3c.south) -- (constrs);
        \draw[->,thick] (topology.south) -- (constrs);

        \draw[->,thick] (constrs) -- (solution);

        \draw[->,thick] (solution) -- (v4a.north);
        \draw[->,thick] (solution) -- (v4b.north);
        \draw[->,thick] (solution) -- (v4c.north);
        \draw[->,thick] (solution) -- (v4s.north);

        \draw[thick] (v2c) -- (dummy);
        \draw[->,thick] (dummy) |- (v5c);

        \draw[->,thick] (v4a) -- (v5a);
        \draw[->,thick] (v4b) -- (v5b);
        \draw[->,thick] (v4c) -- (v5c);
        \draw[->,thick] (v4s) -- (v5s);
\end{tikzpicture}}
\caption{An interface reconciliation workflow performed by the toolchain}
\label{fig:workflow}
\end{minipage}
\end{figure}

To solve the problem, we introduce service structuring that consists of two
parts: a \emph{core} that contains an implementation code as presented in
\cref{fig:service-code}; and a \emph{shell} that is used for (re)routing
messages and renaming labels in output messages.  Such separation facilitates
decontextualisation and service reuse.  The core is generic and does not need
to be changed to match the context; the shell, however, is specific to the
context.  Since the number of input/output channels and label names varies from
context to context, the shell provides a flexible mechanism for service
integration.  In the example in \cref{fig:core-shell}, the core of the service
has one input and one output channel and the environment provides two input and
two output channels.  The shell provides the declaration that messages from
both input channels must be merged into one input channel and output messages
must be copied to two output channels with label \texttt{result} renamed to
\texttt{factorial} and \texttt{square}.  As a result, the output interfaces
that are derived from the services are associated with each of output channels
are $\choice{\eel{factorial}{}{\record{\eel{c}{}{\texttt{int}}}{}}}{}$ and
$\choice{\eel{square}{}{\record{\eel{c}{}{\texttt{int}}}{}}}{}$.  Without the
shell, the interface of the service would be derived using the mechanism from
\cref{sec:protocol}:
$\choice{\eel{result}{}{\record{\eel{c}{}{\texttt{int}}}{}}}{}$.

\label{sec:implementation}

\paragraph{Implementation.} The toolchain is developed in \cpp{} and OCaml,
also using  Clang for the \cpp{} code analysis and the PicoSAT solver as part
of the \cspws{} solver.  The toolchain configures the service interfaces in
five steps as illustrated in \cref{fig:workflow}:
\begin{enumerate}
    \itemsep0em
    \item For each file that contains a service source code in \cpp{}, augment
        them with macros acting as placeholders for the code that enables
        flow inheritance (as illustrated in \cref{fig:annotated-service});
    \item Derive the interfaces from the code (see \cref{fig:service-code});
    \item Given the interfaces and the application topology, construct the
        constraints to be passed on to the \cspws{} algorithm;\label{item:start}
    \item Run the \cspws{} solver;
    \item Based on the solution, generate header files for every service with
        macro definitions.  In addition, the tool generates
        the API functions to be called when a service sends
        or receives a message;\label{item:end}
    \item Given the modified service source files and the generated header
        files, create a binary library for every service.
\end{enumerate}

The steps~\ref{item:start}--\ref{item:end} are performed by a centralised
\emph{composition coordinator} (which could be any service selected
arbitrarily) while other configuration steps can be performed independently for
each service.  The coordinator only receives information about service
interfaces.

As a result, the toolchain generates a set of service libraries from the
topology and service source files.  The libraries can be used with any runtime,
which is able to communicate with services using the API\@.  Application
execution strategy and resource management are up to designers of a particular
web service technology.  In addition to flexibility, advantages of the
presented design are the following:
\begin{itemize}
    \item Interfaces and the code behind them can be generic as long as they
        are sufficiently configurable. No communication between services
        designers is necessary to ensure consistency in the design.
    \item Configuration and compilation of every service is separated from the
        rest of the application.  This prevents source code leaks in
        proprietary software running in the Cloud.
\end{itemize}

\section{Conclusion and Future Work}

We have presented a static interface configuration mechanism of service-based
applications based on CSP and SAT\@.  We rely on the Message Definition
Language that defines a term algebra and is used for specifying flexible
service interfaces.  We defined the format of services written in \cpp{} to
demonstrate the binding between the MDL and message processing functions.  Our
approach supports subtyping, polymorphism and flow inheritance, thanks to the
order relation defined on MDL terms.  We developed a toolchain that
automatically derives service interfaces from the code, generates constraints,
solves the CSP using an original algorithm and propagates the result with
configuration information back to the services.

Currently, our mechanism supports services that respond to at most one message
at a time, avoiding any form of message accumulation.  This is due to the fact
that the linkage between input and output is done using explicit Boolean
selectors rather than from a behavioural description such as the component's
session type.  However, the term type of a component as defined by us could be
the end-result of a derivation based not only on the analysis of the source
code but also on the component's session type, in which case it could be
possible to \emph{derive} the Boolean selectors from it. An integration between
MDL and session types is in our future plans.

Our results may prove useful to the software-as-a-service community, because
our mechanism supports more flexible interfaces than are currently available
without exposing the source code of proprietary software behind them.  Building
services the way we do could enable service providers to configure a solution
for a network customer based on the services that they have at their disposal
as well as those provided by other providers and the customer themselves, all
solely on the basis of interface definitions and automatic tuning to non-local
requirements.

\bibliographystyle{unsrturl}
\bibliography{references}
\end{document}